\RequirePackage[mathlines]{lineno} 
\documentclass[aps,prd,twocolumn,showpacs,amsmath,amssymb]{revtex4-1}
\usepackage{epsfig}
\usepackage{graphicx}
\usepackage{dcolumn}
\usepackage{bm}
\usepackage{overpic}
\usepackage{subfigure}
\usepackage{float}
\usepackage{color}
\usepackage{amsmath}
\usepackage{mathcomp}
\usepackage{multirow}
\usepackage{rotating}
\begin{document}
\normalsize
\parskip=5pt plus 1pt minus 1pt

\title{ \boldmath Search for the rare decay $D^+ \to D^0 e^+\nu_e$ }

\author{
  \begin{small}
    \begin{center}
    M.~Ablikim$^{1}$, M.~N.~Achasov$^{9,d}$, S. ~Ahmed$^{14}$, X.~C.~Ai$^{1}$, O.~Albayrak$^{5}$, M.~Albrecht$^{4}$, D.~J.~Ambrose$^{45}$, A.~Amoroso$^{50A,50C}$, F.~F.~An$^{1}$, Q.~An$^{38,47}$
% ...
    \\
\vspace{0.2cm}
(BESIII Collaboration)\\
\vspace{0.2cm} {\it
$^{1}$ Institute of High Energy Physics, Beijing 100049, People's Republic of China\\
% ...
$^{4}$ Bochum Ruhr-University, D-44780 Bochum, Germany\\
$^{5}$ Carnegie Mellon University, Pittsburgh, Pennsylvania 15213, USA\\
% ...
$^{9}$ G.I. Budker Institute of Nuclear Physics SB RAS (BINP), Novosibirsk 630090, Russia\\
% ...
$^{14}$ Helmholtz Institute Mainz, Johann-Joachim-Becher-Weg 45, D-55099 Mainz, Germany\\
% ...
$^{38}$ State Key Laboratory of Particle Detection and Electronics, Beijing 100049, Hefei 230026, People's Republic of China\\
% ...
$^{45}$ University of Rochester, Rochester, New York 14627, USA\\
% ...
$^{47}$ University of Science and Technology of China, Hefei 230026, People's Republic of China\\
% ...
$^{50}$ (A)University of Turin, I-10125, Turin, Italy; (B)University of Eastern Piedmont, I-15121, Alessandria, Italy; (C)INFN, I-10125, Turin, Italy\\
% ...
$^{d}$ Also at the Novosibirsk State University, Novosibirsk, 630090, Russia\\
% ...
}
    \end{center}
    \vspace{2cm}
  \end{small}
}

\affiliation{}


\vspace{4cm}

\date{\today}

\begin{abstract}
We search for the rare decay $D^+\to D^0 e^+\nu_{e}$, using a data set
with an integrated luminosity of 2.93 $\rm fb^{-1}$ collected at
$\sqrt{s}=3.773$~GeV with the BESIII detector operating at the BEPCII storage rings.
No signals are observed. We set the upper limit on the branching
fraction for $D^+\to D^0 e^+\nu_e$ to be $1.0\times 10^{-4}$ at the 90\%
confidence level.
\end{abstract}


\maketitle

\section{Introduction}
\label{sec:introduction}

Experimental study of the rare decay $D^+\to D^0e^+\nu_e$
is useful to test
standard model predictions~\cite{the1,the2,the3,the4,the5}.
In the semileptonic decay process $D^+\to D^0 e^+\nu_{e}$, the heavy quark flavor
($c$) remains unchanged, and the weak decay proceeds within
the light quark sectors. In the limit of flavor
SU(3) symmetry of the light quarks, the matrix elements of
the weak current can be constrained and the form factors describing
the strong interaction in this decay can be obtained.
Hence, the decay rate of $D^{+}\to D^0e^+\nu_e$ is predicted to be
about $2.78\times10^{-13}$~\cite{the6}.
The experimental potential on this decay at BESIII is discussed in
Ref.~\cite{the6} based on the threshold production of $D^+D^-$ pairs at the $\psi(3770)$ peak.
The reference suggests to search for a neutral $D$ meson in the decay
of $D^+$ when the other $D^-$ in the event is reconstructed
with six tag modes of
$K^+\pi^-\pi^-$, $K^+\pi^-\pi^-\pi^0 $, $K^0_S\pi^-$,
$K^0_S\pi^-\pi^0$, $K^0_S\pi^+\pi^-\pi^-$, and $K^+K^-\pi^-$.
Here, the positron $e^+$ is not required to be reconstructed, since it is very soft in the BESIII detector.

In this paper, the search for  $D^{+}\to D^0e^+\nu_e$ is
carried out using a data set with integrated luminosity of
$2.93~\text{fb}^{-1}$~\cite{lumbesiii} collected at the center-of-mass
energy $\sqrt{s}=3.773$~GeV with the BESIII detector.   At this
energy, $D^+D^-$ pairs are produced without any additional
hadrons. In the analysis, the $D^0$ is
reconstructed through the three decay modes  $K^-\pi^+$,
$K^-\pi^+\pi^+\pi^-$ or $K^-\pi^+\pi^0$, while the tagged $D^-$ is
reconstructed using the six modes as suggested in Ref.~\cite{the6}.
Throughout the paper, charge-conjugate modes are implicitly assumed, unless otherwise noted.

The structure of this paper is as follows. In Sec.~\ref{sec:BESIII},
the BESIII detector and Monte Carlo (MC) simulations are described.
 In Sec.~\ref{sec:selection}, the event selection and the determination of the upper limit on the branching fraction for $D^+\to D^0 e^+\nu_{e}$
are described.
Sec.~\ref{sec:syserr} describes the systematic uncertainties
in the measurement. A short summary of the result is given in Sec.~\ref{sec:summary}.

\section{BESIII detector and MC samples}
\label{sec:BESIII}
The BESIII detector is described in detail elsewhere~\cite{bes3-dector}.
It has an effective geometrical acceptance of 93\% of 4$\pi$.
It consists of a small-cell, helium-based (40\% He, 60\% C$_3$H$_{8}$) main drift chamber (MDC),
a plastic scintillator time-of-flight system (TOF),
a CsI(Tl) electromagnetic calorimeter (EMC) and
a muon system containing resistive plate chambers in the iron return yoke of the 1~T superconducting solenoid.
The momentum resolution for charged tracks is 0.5\% at 1~GeV/$c$.
The photon energy resolution at 1~GeV is 2.5\% in the barrel and 5\% in the endcaps.

A GEANT4-based~\cite{GEANT4-Col,GEANT4-Allison} MC simulation software BOOST~\cite{dengzy},
which includes the geometric description and a simulation of the response of the detector, is used
to determine the detection efficiency and to estimate the potential backgrounds. An `inclusive' MC sample,
which includes generic $\psi(3770)$ decays, initial state radiation (ISR) production of $\psi(3686)$ and
$J/\psi$, QED ($e^+e^-\to e^+e^-,\mu^+\mu^-,\tau^+\tau^-$) and $q\bar{q}~(q=u,d,s)$ continuum process,
is produced at $\sqrt{s}=3.773$ GeV with more than 10 times statistics of data.
The MC events of $\psi(3770)$ decays are produced by a combination
of the MC generators KKMC~\cite{kkmc} and PHOTOS~\cite{poto}, in which the effects of
 ISR~\cite{isr}, final state radiation (FSR) and beam energy spread are considered.
The known decays modes  are generated
using EvtGen~\cite{evtgen} with the branching fractions taken from the Particle Data Group (PDG)~\cite{pdg}.
The remaining unknown decay modes of the charmoinum states are generated using LundCharm~\cite{lundcharm}. The signal MC samples include
a $D^-$ decaying into the six tag modes and a $D^+$ decaying into
$D^0 e^+\nu_e$, where the $D^0$ decays into three specific reconstruction modes.

\section{Event Selection And Data Analysis}
\label{sec:selection}
\vspace{-0.4cm}
Charged tracks are required to be well measured and to satisfy criteria based on the track fit quality;
the angular range is restricted to $|\cos\theta|<0.93$, where $\theta$ is the polar angle with respect to the direction of the positron beam.
Tracks (except for those from $K^0_S$ decays) are also required to have a point of closest approach to the interaction point (IP) satisfying $|V_z| < 10$~cm in the beam
direction and $|V_r| < 1$~cm in the plane perpendicular to the beam direction.
Information from the  $dE/dx$ in the MDC and the flight time obtained from the TOF is used to
identify charged kaons and pions: for each hypothesis $i$, a probability $\mathcal{P}(i)$ is derived, and the probability is required to be ${\cal{P}}(K)> {\cal{P}}(\pi)$, ${\cal{P}}(K)> 0.001$  for kaons and vice-versa for pions.
As suggested in Ref.~\cite{the6}, positrons are not reconstructed since their momentum in the decay $D^+\to D^0 e^+ \nu_{e}$ is less than 5 MeV/$c$.
Electromagnetic showers are reconstructed by clustering hits in the EMC crystals, and the energy resolution is improved by including
the energy deposited in nearby TOF counters.
 To identify photon candidates, showers must have minimum energies of
 25~MeV in the barrel ($|\cos\theta|<0.80$) or 50~MeV in the endcap ($0.86<|\cos\theta|<0.92$). The angle between the shower direction
 and all track extrapolations to the EMC must be larger than $10^{\circ}$.
The time information from the EMC is also required to be in the range 0-700~ns to suppress electronic noise and energy deposits unrelated to the event.
  The $\pi^0$ candidates are selected by requiring the diphoton invariant mass to be within
 $M_{\gamma\gamma}\in (0.110,0.155)$ GeV/$c^2$. Candidates with both photons being detected in the
endcap regions are rejected due to poor
resolution. To improve resolution and reduce background,
the invariant mass of each photon pair is constrained to the nominal $\pi^0$ mass by one-constraint (1C) kinematic fit with the requirement $\chi^2_{\rm 1C}<20$ imposed.
The $K^0_S$ candidates are reconstructed from the combinations of two tracks with opposite charge which satisfy $|\cos\theta|< 0.93$ and $|V_{z}|< 20\,\text{cm}$, but without requirements on $V_{r}$ and particle identification (PID).
The $K^0_S$ candidates must have an invariant mass in the range $0.486<M_{\pi^+\pi^-}<0.510\;\text{GeV}/c^2$.
To suppress the random combinational backgrounds and reject the mis-combinations of pion pairs, the ratio of the
flight distance of $K^0_S$ ($L$) over its uncertainty ($\sigma_{L}$), $L/\sigma_{L}$, is required to be larger than 2.

 The single tag (ST) $D^{-}$ candidate events are selected by reconstructing a $D^{-}$ in
the following hadronic final states: $K^+\pi^-\pi^-$, $K^+\pi^-\pi^-\pi^0 $, $K^0_S\pi^-$,
$K^0_S\pi^-\pi^0$, $K^0_S\pi^+\pi^-\pi^-$, and $K^+K^-\pi^-$, comprising approximately 28.0\% ~\cite{pdg} of all $D^-$ decays.

To identify the reconstructed $D^{-}$ candidates in the tag modes, we use two variables: the beam energy constrained mass, $M_{\rm BC}$, and the energy
difference, $\Delta E$, which are defined as
\begin{eqnarray}
\begin{aligned}
M_{\rm BC} \equiv \sqrt{E^2_{\rm beam}/c^4-|\vec{p}_{D^{-}}|^2/c^2}, \Delta E \equiv {E_{D^{-}}} - E_{\rm beam},
\end{aligned}
\end{eqnarray}
where $\vec{p}_{D^-}$ and $E_{D^{-}}$ are the reconstructed momentum and energy of the $D^{-}$ candidate in
the $e^+e^-$ center-of-mass system, and $E_{\rm beam}$ is the beam energy. For the true $D^{-}$ candidates,
$\Delta E$ is consistent with zero, and $M_{\rm BC}$ is consistent with the $D^{-}$ mass.
We accept $D^-$ candidates with $M_{\rm BC}$ greater than 1.83 GeV/c$^2$ and with mode-dependent
$\Delta E$ requirements of approximately three standard deviations around the $\Delta E$ peaks.
For the ST modes, we accept at most one candidate per mode per event if there are multi-candidates; the candidate one with the smallest $|\Delta E|$
is chosen~\cite{deltaE}.

\begin{figure}[hbtp]
\centering
\epsfig{width=0.485\textwidth,clip=true,file=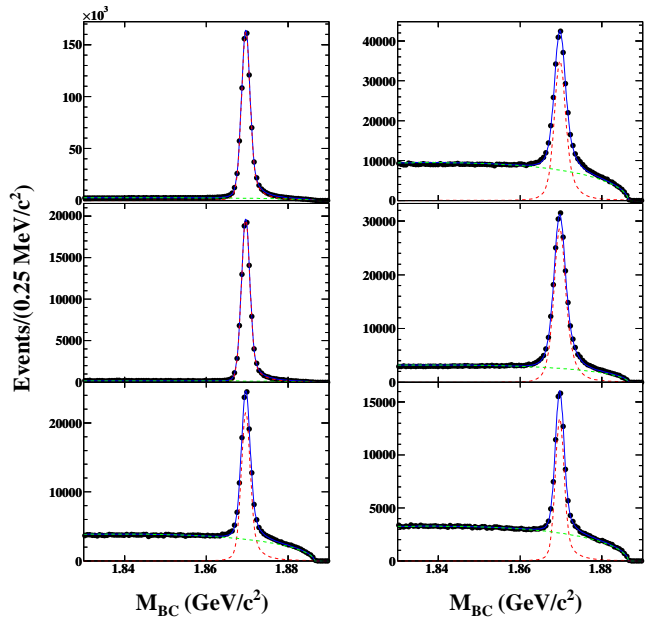}
\caption{(Color online) Fits to the $M_{\rm BC}$ distributions of the ST modes of (a) $K^+\pi^-\pi^-$, (b) $K^+\pi^-\pi^-\pi^0$, (c) $K^0_S\pi^-$,
 (d) $K^0_S\pi^-\pi^0$, (e) $K^0_S\pi^-\pi^+\pi^-$ and (f) $K^+K^-\pi^-$. Data are shown as points,
the blue solid lines are the total fits, the green dashed lines are the background shapes,
and the red dotted lines are the signal shapes.}
\label{mbcfit}
\end{figure}

To obtain the ST yields, we fit the $M_{\rm BC}$ distributions of the accepted $D^-$ candidates, as shown in Fig.~\ref{mbcfit}.
The signal shape is modeled by a MC-determined shape convoluted with a Gaussian function.
The signal line shape includes the effects of beam energy spread, ISR, the $\psi(3770)$ line shape, and detector resolution.
Combinatorial background is modeled by an ARGUS function~\cite{argus}.
The tag efficiency is studied using inclusive MC samples following
the same procedure. The $\Delta E$ requirements, ST yields in data and the corresponding ST efficiencies are listed in Table~\ref{tab::steff}.
The total ST yield is $N^{\rm tot}_{\rm ST} = 1555039\pm1471$ events.

\begin{table}[hbtp]
\begin{center}
\caption{The summary of $\Delta E$ requirements, ST yields in data ($N_{\rm ST}^{j}$) and ST efficiencies ($\epsilon_{\rm ST}^j$). Branching fractions
of the $K^0_S$ and $\pi^0$ decays are not included in the efficiencies.
$j$ denotes the ST mode. The uncertainties are statistical only.}
\ \\
\begin{tabular}{lcrcc}
\hline
\hline
Mode $j$                       & $\Delta E$ (MeV)    &\multicolumn{1}{c}{$N_{\rm ST}^j$}  & $\epsilon_{\rm ST}^j$(\%)\\
\hline
$K^+\pi^-\pi^-$            & $(-30,30)$    &  $826795\pm973$  & $53.23\pm0.02$\\
$K^+\pi^-\pi^-\pi^0$       & $(-52,39)$    &  $241618\pm696$  & $24.83\pm0.02$\\
$K^0_{S}\pi^-$             & $(-32,32)$     &  $96306\pm324$  & $53.11\pm0.05$\\
$K^0_{S}\pi^-\pi^0$        & $(-57,40)$    &  $203358\pm555$  & $26.02\pm0.02$\\
$K^0_{S}\pi^-\pi^+\pi^-$   & $(-34,34)$    &  $115223\pm436$  & $28.93\pm0.03$\\
$K^+K^-\pi^-$               & $(-30,30)$    &  $71739\pm360$  & $42.61\pm0.05$\\
\hline
\hline
\end{tabular}
\label{tab::steff}
\end{center}
\end{table}

On the recoil side of the $D^-$ mesons, we search for the rare decay $D^+\to D^0 e^+\nu_e$, in which the $D^0$ meson is reconstructed using $D^0\to K^-\pi^+$, $K^-\pi^+\pi^+\pi^-$, and $K^-\pi^+\pi^0$. If a $D^0$ meson can be found, we label the events to be a double tag (DT) event.

\begin{figure}[hbtp]
\centering
\epsfig{width=0.485\textwidth,clip=true,file=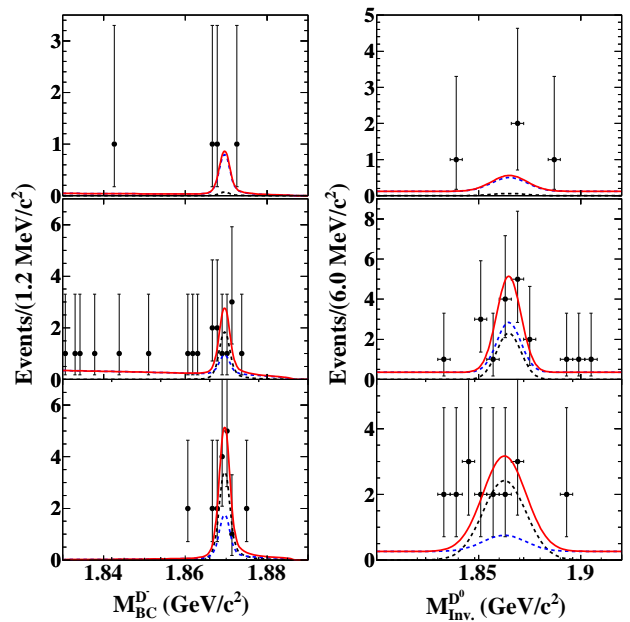}
\caption{(Color online) Projections of the 2D fits to the distributions of $M^{D^-}_{\rm BC}$ (left column) and $M^{D^0}_{\rm Inv.}$ (right column) of the candidates in data with the signal modes (a) $D^0\to K^-\pi^+$, (b) $D^0\to K^-\pi^+\pi^+\pi^-$ and (c) $D^0\to K^-\pi^+\pi^0$.
The dots with error bars are data, the red solid lines show the fit results, the black dashed lines represent the signal shapes, and the blue dotted lines represent total background shapes.}
\label{2dfitdata}
\end{figure}

With the DT technique, the continuum background $e^{+}e^{-}\to q\bar{q}$ is highly suppressed. The remaining background dominantly comes from $D\bar{D}$ events with a correctly reconstructed signal $D^0$ or tag $D^{-}$ while the opposite side is misreconstructed.
These background can be suppressed by studying the two uncorrelated variables, $D^0$ momentum and observed $D^-D^0$ energy distributions in signal MC and inclusive MC simulation.
A probability is constructed by multiplying the normalized $D^0$ momentum distribution and the normalized observed $D^-D^0$ energy distribution.
To obtain reliable event selection criteria and improve the ratio of signal over background, an optimization
is performed using the inclusive MC samples, in which the branching fraction
of this rare decay is set to be $10^{-6}$ that is predicted in Ref.~\cite{the6}. The background yields from the inclusive MC samples are obtained from two-dimensional (2D) fits to the
beam-energy constrained mass for the $D^-$ candidates ($M^{D^-}_{\rm BC}$) and the distributions of the invariant mass for the $D^0$ candidates ($M^{D^0}_{\rm Inv.}$).
In the 2D fits, the signal shape of $M^{D^-}_{\rm BC}$ is modeled using a MC-determined shape
and the background shape is modeled with an ARGUS function~\cite{argus};
the signal shape of $M^{D^0}_{\rm Inv.}$ is modeled using a Gaussian function and the
background shape is modeled with a polynomial function.
Based on the optimization,
the probability is required to be larger than 0.37, 0.34, and 0.54 for the signal modes $D^0\to K^-\pi^+$, $K^-\pi^+\pi^+\pi^-$,
and $K^-\pi^+\pi^0$, respectively. The events satisfying these requirements are kept for further analysis. The DT efficiencies for the individual tag modes and $D^0$ reconstruction modes, as well as the ST yield weighted efficiencies of reconstructing $D^+\to D^0 e^{+} \nu_{e}$ are listed
in Table II.
2D fits are performed on the accepted events for each signal mode in data, as shown in Fig.~\ref{2dfitdata}. We obtain the fit
yields $N_{\rm data}^{\rm obs.}$ to be $0.2\pm2.8$, $5.9\pm2.9$, and $10.0\pm4.3$ for the signal modes
$D^0\to K^-\pi^+$, $D^0\to K^-\pi^+\pi^+\pi^-$, and $D^0\to K^-\pi^+\pi^0$, respectively.
In the fit, the analogous functions as those fits to the inclusive MC sample are imposed.
To consider the detector resolution difference between data and MC simulation, the $M^{D^-}_{\rm BC}$ signal shape is convoluted with a Gaussian function with parameters obtained by fitting the $M_{\rm BC}^{D^-}$ distribution of the ST candidate events and the $M^{D^0}_{\rm Inv.}$ signal shape is convoluted with another Gaussian function with parameters determined by studying the associated DT hadronic $D^0\bar D^0$ events.

\begin{table}[hbtp]
\begin{center}
\caption{ The DT efficiencies ($\epsilon^{\rm DT}_{ji}$) and the efficiency of reconstructing $D^+\to D^0e^+\nu_e$ weighted by
   the ST yields ($\epsilon^{i}$),
 where $j$ denotes the ST mode and $i$ denotes the signal mode. Branching fractions of the $K^0_S$ and $\pi^0$ decays are not
    included in the efficiencies. The uncertainties are statistical only.}
\begin{tabular}{cccc}
\hline\hline
Mode              &$K^-\pi^+$ (\%)  &$K^-\pi^+\pi^+\pi^-$ (\%) &$K^-\pi^+\pi^0$ (\%)\\
\hline
$K^+\pi^-\pi^-$         &19.43 $\pm$ 0.13  &11.69 $\pm$ 0.10    &6.39 $\pm$ 0.08  \\
$K^+\pi^-\pi^-\pi^0$    &8.91  $\pm$ 0.09  &4.79  $\pm$ 0.07    &3.17 $\pm$ 0.06  \\ 	
$K^0_{S}\pi^-$          &20.06 $\pm$ 0.13  &11.68 $\pm$ 0.10	&6.51 $\pm$ 0.08  \\
$K^0_{S}\pi^-\pi^0$     &9.90 $\pm$ 0.09  &5.27  $\pm$ 0.07	    &3.24 $\pm$ 0.06  \\
$K^0_{S}\pi^+\pi^-\pi^-$&10.49 $\pm$ 0.10  &5.45  $\pm$ 0.07	&3.18 $\pm$ 0.06  \\
$K^+K^-\pi^-$           &14.77 $\pm$ 0.11  &8.83  $\pm$ 0.09	&5.06 $\pm$ 0.07  \\\hline
$\epsilon^{i}$          &36.42 $\pm$ 0.07&20.95 $\pm$ 0.06& 12.06 $\pm$ 0.04\\
\hline\hline
\end{tabular}
\end{center}
\label{tab::dteff}
\end{table}

Peaking backgrounds are obtained by fitting the distributions of inclusive MC samples as done in the optimization process.
The normalized background numbers $N^{i}_{\rm bkg}$ are obtained to be $2.8\pm0.6$, $6.0\pm0.9$, and $12.4\pm1.3$ for
the signal modes $D^0\to K^-\pi^+$, $D^0\to K^-\pi^+\pi^+\pi^-$, and $D^0\to K^-\pi^+\pi^0$, respectively.
And all the backgrounds arise from the $D^0\bar{D}^0$ and $D^{+}D^{-}$ events. The uncertainties in $N^i_{\rm bkg}$ are dominated by the limited MC sample size, and the uncertainties of the luminosity of data, the $D^0\bar D^0(D^+D^-)$ cross sections, the quoted branching fractions of $D^{0(+)}$ decays and the data-MC difference of the efficiencies of the $K^+(\pi^+)$ tracking (PID) and the $\pi^0$ reconstruction can be negligible.

The expected signal yield in a specific signal mode ($N_{\rm sig}^{i}$) can be expressed as
\begin{equation}
N_{\rm sig}^{i} = N^{\rm tot}_{\rm ST} \times \epsilon^{i} \times {\cal{B}}^{i} \times {\cal{B}}_{D^{+}},
\label{bran1}
\end{equation}
where $i = 0,1,2$, represent the signal modes $D^0\to K^-\pi^+$, $K^-\pi^+\pi^+\pi^-$, and $K^-\pi^+\pi^0$, respectively; $N^{\rm tot}_{\rm ST}$ represents the total ST yield in data;
$\epsilon^{i}$ represents the efficiency of reconstructing $D^+\to D^0e^+\nu_e$ for the signal mode $i$, which is weighted by the ST yields;
${\cal{B}}^{i}$ represents the quoted branching fraction of $D^0\to K^-\pi^+$, $K^-\pi^+\pi^+\pi^-$ or $K^-\pi^+\pi^0$ quoted from the PDG~\cite{pdg}; ${\cal{B}}_{D^{+}}$ is the branching fraction of $D^+\to D^0 e^+ \nu_{e}$.

The expected signal yield can also be expressed as
\begin{equation}
N_{\rm sig}^{i} = N^{{\rm obs.}i}_{\rm data} - N^{i}_{\rm bkg },
\label{bran2}
\end{equation}
where $N^{{\rm obs.}i}_{\rm data}$ represents the number of events from the 2D fit in data,
$N^{i}_{\rm bkg }$ represents the expected background event number estimated by fitting the inclusive MC sample.

Since there is no obvious signal observed in data, an upper limit on the branching fraction of $D^+\to D^0 e^+\nu_{e}$ is determined.
For each signal mode, the likelihood value is obtained by treating ${\cal{B}}_{D^+}$ as a free parameter in the Eq.~(\ref{bran1}).  The resulting likelihood function is labeled as $\mathcal L_i$. To combine the three $D^0$ signal modes,
 a joint likelihood function is constructed by
${\cal{L}}_{\rm com} = \mathcal L_1\times \mathcal L_2\times\mathcal L_3$.
Based on the Bayesian method,
the upper limit on the branching fraction for $D^+\to D^0 e^+\nu_e$ is determined to be
${\mathcal B}(D^+\to D^0 e^+\nu_e)<9.0\times 10^{-5}$ at the 90\% confidence level, by integrating $\mathcal L_{\rm com}$
from 0 up to 90\% of the area in the physical region.

\section{Systematic Uncertainties}
\label{sec:syserr}
The sources of systematic uncertainty considered in the determination of the upper limit on ${\cal{B}}(D^+\to D^0 e^+\nu_{e})$
  are listed in Table~\ref{table:syserrA} and
  described below.

\begin{itemize}
 \item {\bf Signal side:}
 The systematic uncertainties in the ST selection cancel.
 Concerning the signal side, the systematic uncertainties in the
  tracking and PID efficiencies, $\pi^0$ reconstruction efficiency, as well as the quoted branching
  fractions are
  assigned relative to the measured branching fraction.
 \begin{itemize}
 \item \noindent\textbf{Tracking and PID efficiency:} The tracking and PID efficiencies of $K^+$ and $\pi^+$ are investigated by
using DT $D\bar{D}$ hadronic events. The difference of the tracking and PID efficiencies between data and MC simulation
is assigned as 1\% per track, individually.
 \item \noindent\textbf{{\bf$\pi^0$} reconstruction:} The $\pi^0$ reconstruction efficiency is studied by examining
the DT hadronic decays $D^0\to K^-\pi^+$ and $K^-\pi^+\pi^+\pi^-$ versus $\bar{D}^0\to K^-\pi^+\pi^0$ and
$K^0_S(\pi^+\pi^-)\pi^0$. The difference of the $\pi^0$ reconstruction efficiency between data and MC simulation is
estimated to be 2\% per $\pi^0$.
 \item \noindent\textbf{Quoted branching fractions:}
The uncertainties of the quoted branching fractions are 1.0\%, 2.9\%, and 5.6\% for $D^0\to K^-\pi^+$, $K^-\pi^+\pi^+\pi^-$, and $K^-\pi^+\pi^0$, respectively~\cite{pdg}.
\end{itemize}

The quadratic sums of the systematic uncertainties from {\bf Signal side} are 3.0\%, 6.4\%, and 6.6\% for  $D^0\to K^-\pi^+$, $K^-\pi^+\pi^+\pi^-$, and $K^-\pi^+\pi^0$, respectively.
The combined uncertainty on the branching fraction from {\bf Signal
  side} is estimated by convoluting the likelihood distribution with a
Gaussian function representing the systematic uncertainty, and
the relative change of the upper limit on ${\cal{B}}(D^+\to D^0 e^+\nu_{e})$, 3.3\%, is taken as a systematic uncertainty.
 \item {\bf Background estimation:}
The systematic uncertainty associated with the background estimation
is studied by changing the background yield $N^i_{\rm bkg}$ by 1 standard deviation.
The relative change of the upper limit on ${\cal{B}}(D^+\to D^0 e^+\nu_{e})$, 13.3\%, is taken as a systematic uncertainty.
 \item \noindent\textbf{MC statistics:} Detailed studies show that the upper limit on $\mathcal B(D^+\to D^0e^+\nu_e)$ is insensitive to the uncertainties due to the limited MC statistics (0.5\%). So, they are negligible in this analysis.
 \item \noindent\textbf{$M_{\rm BC}$ fit (ST):} The systematic uncertainty associated with the ST yields extracted by fitting $M_{\rm BC}$ distribution is estimated to be 0.5\% by varying the fit range, signal shape and endpoint of the ARGUS function. The variation of the upper limit on $\mathcal B(D^+\to D^0e^+\nu_e)$ arising from different $M_{\rm BC}$ fits is found to be negligible.
 \item \noindent\textbf{Probability requirement:} The systematic uncertainty in the probability requirement is investigated
by changing the requirement by $\pm0.01$. The effect on the upper limit of ${\cal{B}}(D^+\to D^0 e^+\nu_{e})$, 2.3\%, is taken as a systematic uncertainty.

\item \noindent\textbf{2D fit:}
  The systematic uncertainty of the 2D fit to the DT candidates is investigated by varying the parameters of the smeared Gaussian functions by 1 standard deviation. The impact on the upper limit of $\mathcal B(D^+\to D^0e^+\nu_e)$, 2.5\%, is taken as a systematic uncertainty.
\end{itemize}

Assuming that all systematic uncertainties are independent, we add
them in quadrature and obtain a total systematic uncertainty of 14.4\%

The final upper limit on $\mathcal B(D^+\to D^0e^+\nu_e)$ is determined
by incorporating the systematic uncertainty. Here, the systematic uncertainty is considered
by convoluting
the likelihood distribution with a Gaussian function with a relative width of 14.4\%. The resulting upper limit on $\mathcal B(D^+\to D^0e^+\nu_e)$ is estimated to be $1.0\times 10^{-4}$ at the 90\% confidence level.

\begin{table*}[hbtp]
\begin{center}
\caption{Summary of the relative systematic uncertainties (in \%), where the $2^{\rm nd}$-$5^{\rm th}$ rows are assigned relative to the measured branching fraction, while the others are assigned by the effects on the upper limit of ${\cal{B}}(D^+\to D^0 e^+\nu_{e})$}.
\ \\
\begin{tabular}{c|ccc}
\hline
\hline
Source  & $D^0\to K^-\pi^+$ & $D^0\to K^-\pi^+\pi^+\pi^-$ & $D^0\to K^-\pi^+\pi^0 $ \\
\hline
Tracking         &  2.0 & 4.0 & 2.0  \\
PID              &  2.0 & 4.0 & 2.0 \\
Quoted branching fraction                 &  1.0 & 2.9 & 5.6  \\
$\pi^0$ reconstruction         &  -   & - & 2.0  \\
\hline
Sum of {\bf Signal side}     &  3.0 & 6.4 & 6.6 \\\hline
{\bf Signal side}                &  \multicolumn{3}{c}{4.4} \\
Background estimation                &  \multicolumn{3}{c}{13.3} \\
MC statistics              &  \multicolumn{3}{c}{negligible} \\
$M_{\rm BC}$ fit (ST)           &  \multicolumn{3}{c}{negligible} \\
Probability requirement               &  \multicolumn{3}{c}{2.3} \\
2D fit               &  \multicolumn{3}{c}{2.5} \\\hline
Total                      &  \multicolumn{3}{c}{14.4} \\
\hline
\hline
\end{tabular}
\label{table:syserrA}
\end{center}
\end{table*}

\section{Summary}
\label{sec:summary}

In summary, we perform a search for the rare decay $D^+\to D^0 e^+
\nu_{e}$, using 2.93~fb$^{-1}$ data taken at $\sqrt{s}= 3.773$~GeV
with the BESIII detector operating at the BEPCII collider. A double
tag method is used, without reconstructing the electron in the final state.
No obvious signal is observed, and the upper limit on the branching fraction
for $D^+\to D^0 e^+ \nu_{e}$ is estimated to
be $1.0\times10^{-4}$ at the 90\% confidence level.
Due to the limited data sample, the measured upper limit is far above the theoretical prediction by Ref.~\cite{the6}.
As the first search for the $D^+\to D^0 e^+\nu_e$, however, it provides complementary experimental information for the
understanding of the SU(3) flavor symmetry in $D$ decays~\cite{su3} and the standard model predictions for rare semileptonic decays.

\begin{acknowledgements}
\label{sec:acknowledgement}
The BESIII collaboration thanks the staff of BEPCII and the IHEP computing center for their strong support. This work is supported in part by National Key Basic Research Program of China under Contract No. 2015CB856700; National Natural Science Foundation of China (NSFC) under Contracts Nos. 11235011, 11322544, 11335008, 11425524, 11475055, 11635010, 11605042; the Chinese Academy of Sciences (CAS) Large-Scale Scientific Facility Program; the CAS Center for Excellence in Particle Physics (CCEPP); the Collaborative Innovation Center for Particles and Interactions (CICPI); Joint Large-Scale Scientific Facility Funds of the NSFC and CAS under Contracts Nos. U1232201, U1332201, U1532257, U1532258, U1632109; CAS under Contracts Nos. KJCX2-YW-N29, KJCX2-YW-N45; 100 Talents Program of CAS; National 1000 Talents Program of China; INPAC and Shanghai Key Laboratory for Particle Physics and Cosmology; German Research Foundation DFG under Contracts Nos. Collaborative Research Center CRC 1044, FOR 2359; Istituto Nazionale di Fisica Nucleare, Italy; Koninklijke Nederlandse Akademie van Wetenschappen (KNAW) under Contract No. 530-4CDP03; Ministry of Development of Turkey under Contract No. DPT2006K-120470; The Swedish Research Council; U. S. Department of Energy under Contracts Nos. DE-FG02-05ER41374, DE-SC-0010118, DE-SC-0010504, DE-SC-0012069; U.S. National Science Foundation; University of Groningen (RuG) and the Helmholtzzentrum fuer Schwerionenforschung GmbH (GSI), Darmstadt; WCU Program of National Research Foundation of Korea under Contract No. R32-2008-000-10155-0.

\end{acknowledgements}

\end{document}